\documentclass{ws-procs9x6}

\begin{document}

\title{Acceleration of cosmic rays at supernova remnant shocks: constraints from gamma-ray observations}

\author{M. Lemoine-Goumard}

\address{Centre d'\'Etudes Nucl\'eaires de Bordeaux Gradignan\\
Universit\'e Bordeaux 1, CNRS/IN2P3\\
33175 Gradignan, France\\
E-mail: lemoine@cenbg.in2p3.fr\\
Funded by contract ERC-StG-259391 from the European Community
}

\begin{abstract}
In the past few years, gamma-ray astronomy has entered a golden age. At TeV energies, only a handful
of sources were known a decade ago, but the current generation of ground-based imaging
atmospheric Cherenkov telescopes has increased this number to more than one hundred. At GeV
energies, the Fermi Gamma-ray Space Telescope has increased the number of known sources by
nearly an order of magnitude in its first 2 years of operation. The recent detection and
unprecedented morphological studies of gamma-ray emission from shell-type supernova remnants
is of great interest, as these analyses are directly linked to the long standing issue of the origin of the
cosmic-rays. However, these detections still do not constitute a conclusive proof that supernova remnants 
accelerate the bulk of Galactic cosmic-rays, mainly due to the difficulty of disentangling the hadronic
and leptonic contributions to the observed gamma-ray emission. In this talk, I will review the most
relevant cosmic ray related results of gamma ray astronomy concerning supernova remnants.
\end{abstract}


\bodymatter

\section{The cosmic-ray mystery}\label{intro}

\subsection{The link between cosmic-rays and supernova remnants}
The association between supernova remnants (SNRs) and Galactic cosmic rays (CRs) is very popular since 1934, 
when Baade and Zwicky argued that this class of astrophysical objects can account for the required CR energetics~[\refcite{baade}]. 
Indeed, in order to maintain the cosmic-ray energy density in the Galaxy, about 3 supernovae per century should transform 10 percent of their kinetic energy in cosmic-ray energy. This argument has also been supported by E. Fermi's proposal of a very general mechanism for particle acceleration, which is very efficient if applied at SNR shocks~[\refcite{bell}]. The extremely interesting point of the diffusive shock acceleration (DSA) mechanism is that it naturally yields power-law spectra for the energy distribution of accelerated particles. However, until recently there were absolutely no observational evidence concerning the 
acceleration of protons and nuclei in SNRs. Indeed, through their interaction with the interstellar magnetic fields, the charged particles arriving on Earth have lost all directional information and cannot be used to pinpoint the sources. That is why, almost 100 years after their discovery by V. Hess, the origins of the cosmic-rays and their cosmic accelerators remain
unknown. \\

Astronomy with gamma-rays provides a means to study these sources of high energy particles. Indeed,
cosmic rays (ionized nuclei of all species, but mostly protons, plus a small fraction of electrons) can interact with ambient matter and photons producing
gamma-rays via two different channels. One mechanism invokes the interaction of accelerated protons at supernova remnants shocks with interstellar material generating neutral pions which in turn decay into gamma rays. We call this mechanism the hadronic scenario. A second competing channel exists in the inverse Compton scattering of the photon fields in the surroundings of the SNR by the same relativistic electrons that generate the synchrotron X-ray emission. This is the leptonic scenario. Being of leptonic or hadronic origin, these gamma-rays are not affected while they travel to Earth and can therefore be used to pinpoint the cosmic accelerators in our Galaxy.

\subsection{Gamma-ray experiments}
Two major breakthroughs in gamma-ray astronomy occurred in recently.\\ Firstly, after more
than 20 years of development, the first source of very high energy gamma-rays, the Crab Nebula, was
discovered in 1989 by the Whipple telescope. Since this date the technical progresses in this field have led to
important scientific results, especially by the Cherenkov telescopes H.E.S.S., VERITAS and MAGIC. These ground-based experiments for gamma-ray astronomy rely
on the development of cascades (air-showers) initiated by astrophysical gamma-rays. Such cascades only persist to ground-level above 1 TeV and only produce significant Cherenkov light above a few GeV, setting a fundamental threshold to the range
of this technique. Today, more than 120 gamma-ray sources have been detected with high significance, 17 being associated to supernova remnants or molecular clouds. \\
Second, in space, the Large Area Telescope (LAT) onboard the Fermi satellite has considerably improved our knowledge of the 0.1-100 GeV gamma-ray sky with 1873 objects detected in only two years of observation [\refcite{2FGLcat}]. It has moved the field from the detection of a small number of sources to the detailed study of several classes of Galactic and extragalactic objects. A complete study of association of the 1873 sources detected show that $\sim 4$\% of them are associated to supernova remnants  [\refcite{2FGLcat}].\\
There is no doubt today that supernova remnants can accelerate efficiently particles up to $10^{14}$~eV. The question is whether these particles are protons or electrons and if they can be accelerated up to the knee of the cosmic-ray spectrum ($10^{15}$~eV).

\subsection{First evidence of efficient particle acceleration in supernova remnants with X-ray satellites}
Accelerated electrons producing gamma-ray emission through inverse Compton scattering also radiate through synchrotron
emission when spiraling in a magnetic field. This emission extends from the radio to the X-ray domain. While radio synchrotron emission is observed in most SNRs (in 203 over the 217 observed Galactic SNRs, [\refcite{green}]), X-ray synchrotron emission is observed only in a few remnants up to now. In some of these X-ray detected SNRs, the X-ray synchrotron
emission exhibits a filamentary emission just behind the blast wave. One plausible explanation is that the magnetic field is
large enough ($\sim 100 \, \mu$G) to induce strong radiative losses in the high energy electrons [\refcite{vink}, \refcite{ballet}]. If the magnetic field is indeed amplified at the limbs, the maximum energy at which particles can be accelerated is much larger there ($>$ 1000 TeV) than outside the limbs (E $\approx$ 25 TeV if B $\approx 10 \, \mu$G).\\
Recently, a discovery of the brightening and decay of X-ray hot spots in the shell of the SNR RX J1713.7-3946 on a one-year timescale has been reported by Uchiyama and collaborators [\refcite{uchiyama}]. This rapid variability implies that electron acceleration needs to take place in a strongly magnetized environment, indicating amplification of the magnetic field by a factor of more than 100.\\
A last evidence of very efficient particle acceleration in supernova remnants is provided by the postshock plasma temperatures observed in SNRs 1E 0102.2-7219 and RCW 86, that are
lower than expected for their measured shock velocities [\refcite{hughes}, \refcite{helder}]. For the first time, by comparing the measured post-shock proton temperature with the one determined using the shock velocity, the authors presented the evidence that $>50$\% of the post-shock pressure is produced by cosmic rays.\\
There are strong indirect arguments confirming that electrons and protons are accelerated up to at
least TeV energies (maybe even PeV) in supernova remnants. A direct signature of accelerated protons is
expected through pion decay emission in the GeV-TeV gamma ray range.

\section{Detection of supernova remnants in gamma-rays}
The sample of supernova remnants detected in gamma-rays is now extremely large: it goes from evolved supernova remnants interacting with molecular clouds (MC) up to young shell-type supernova remnants and historical supernova remnants. The Fermi-LAT even detected one evolved supernova remnants without MC interaction, Cygnus loop. This section will review the main characteristics of the detected SNRs.

\subsection{Supernova remnants interacting with molecular clouds}
\begin{figure}
\begin{center}
\psfig{file=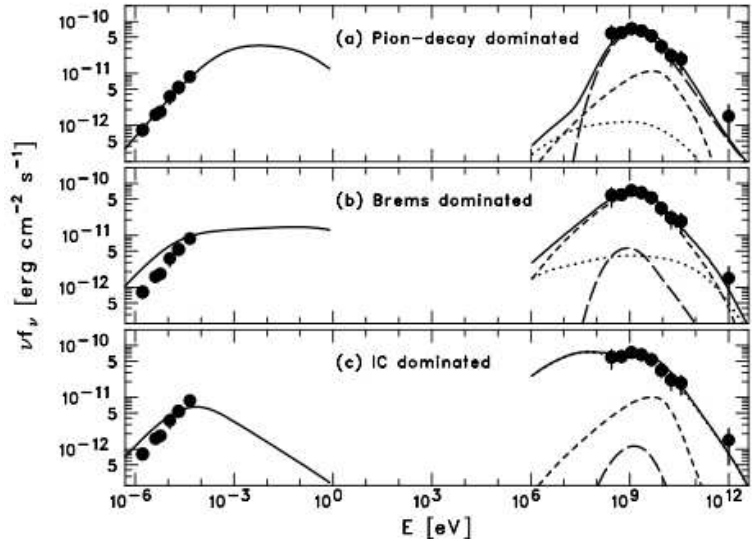,width=4in}
\end{center}
\caption{Different scenarios proposed for the multiwavelength modeling of W51C~[\refcite{fermi_w51}]. The
radio emission (from Moon \& Koo 1994) is explained by synchrotron radiation, while the
gamma-ray emission is modeled by different combinations of pion-decay (long-dashed curve),
bremsstrahlung (dashed curve), and IC scattering (dotted curve). The sum of the three
component is shown as a solid curve. See [\refcite{fermi_w51}] for more details.
}
\label{fig1}
\end{figure}

The Fermi LAT Collaboration has so far reported the discoveries of five middle aged ($\sim 10^4$ yrs) remnants interacting with molecular clouds: W51C~[\refcite{fermi_w51}], W44~[\refcite{fermi_w44}], IC 443~[\refcite{fermi_ic443}], W49~[\refcite{fermi_w49b}] and W28~[\refcite{fermi_w28}]. Apart from W44, they have all been detected in the TeV regime as well. These SNRs are generally much brighter in GeV than in TeV in terms of energy flux (due to a spectral steepening arising at a few GeV), which emphasizes the importance of the GeV observations. The interaction with a molecular cloud provides the target material that allows one to enhance the gamma-ray emission, either through bremsstrahlung by relativistic electrons or by pion-decay gamma-rays produced by high-energy protons. The observed large luminosity
of the GeV gamma-ray emission precludes the inverse-Compton scattering off the CMB and interstellar
radiation fields as the main emission mechanism since it would require an extremely low density (to suppress the bremsstrahlung and proton-proton interaction), a low magnetic field to enhance the gamma/X-ray flux ratio and an unrealistically large energy injected into protons. In addition, the break in the electron spectrum corresponding to the gamma-ray spectrum directly appears in the radio data leading to a bad modeling of the radio data and therefore disfavours the bremsstrahlung process. A model in which gamma-rays are produced via proton-proton interaction gives the most satisfactory explanation for the GeV gamma-rays observed in SNRs interacting with molecular gas as seen in Figure~\ref{fig1} for the case of W51C.\\
There are two different types of hadronic scenarios to explain the GeV gamma-ray emission arising from such SNRs: the "Runaway CR" model [\refcite{aha_escape}, \refcite{ohira}] and the "Crushed Cloud" model [\refcite{uchiyama_CR}]. The Runaway CR model considers gamma-ray emission from molecular clouds illuminated by runaway CRs that have escaped from their accelerators, whereas the Crushed Cloud model invokes a shocked molecular cloud into which cosmic-ray particles are adiabatically compressed and accelerated resulting in enhanced synchrotron and pion-decay gamma-ray emissions.

\subsection{Young shell-type supernova remnants}
\begin{figure}
\begin{center}
\psfig{file=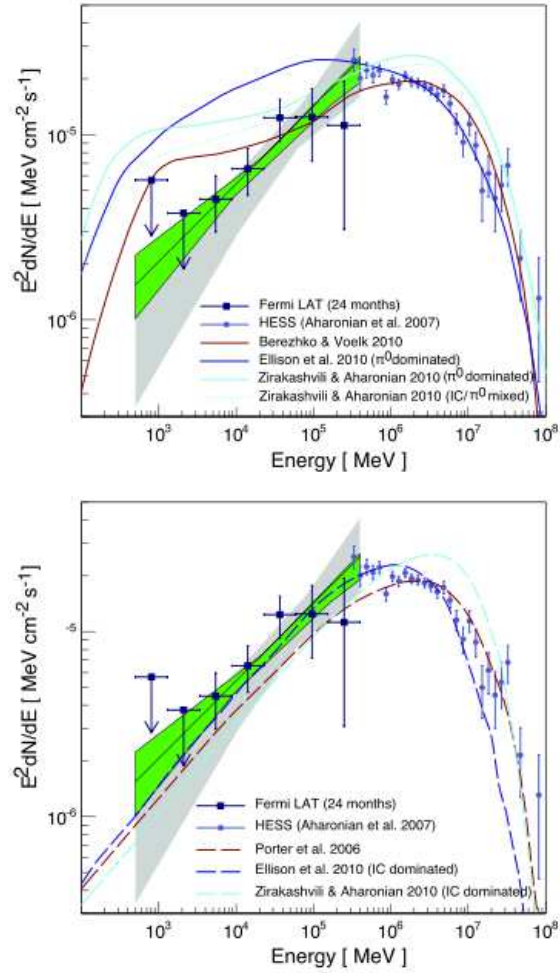,width=3in}
\end{center}
\caption{Energy spectrum of RX J1713.7-3946 in gamma rays. Shown is the Fermi-LAT~[\refcite{fermi_rxj}] detected emission
in combination with the energy spectrum detected by H.E.S.S. [\refcite{aharonian_rxj2}]. See [\refcite{fermi_rxj}]  for more details.
}
\label{fig2}
\end{figure}

Four young shell-like SNRs with clear shell-type morphology resolved in VHE gamma-rays have been detected by H.E.S.S.: RX J1713.7-3946 [\refcite{aharonian_rxj1}, \refcite{aharonian_rxj2}], RX J0852.04622 - also known as Vela Junior - [\refcite{aharonian_velajr1}],
SN 1006 [\refcite{acero_sn1006}] and HESS J1731-347 [\refcite{acero_1731}]. A fifth case, RCW 86 [\refcite{aharonian_rcw86}], might be added to this list although the TeV shell morphology has not yet been clearly proved. Two of them, RX~J1713.7-946 [\refcite{fermi_rxj}] and Vela Junior [\refcite{fermi_velajr}], have been detected by Fermi-LAT allowing direct investigation of young shell-type SNRs as sources of cosmic rays. Concerning RX~J1713.7-3946, the Fermi-LAT spectrum is well described by a very hard power-law with a photon index of $\Gamma = 1.5 \pm 0.1$ that coincides in normalization with the steeper H.E.S.S.-detected gamma-ray spectrum at higher energies. The GeV measurements with Fermi-LAT do not agree with the expected fluxes around 1 GeV in most hadronic models published so far (e.g., Berezhko \& Voelk 2010 [\refcite{berezhko}]) and requires an unrealistically large density of the medium. The agreement with the expected IC spectrum is better (as can be seen in Figure \ref{fig2}) but requires a very low magnetic field of $\sim 10 \, \mu$G in comparison to the one measured in the thin filaments by X-ray observations. It is possible to reconcile a high magnetic field with the leptonic model if GeV gamma rays are radiated not only from the filamentary structures seen by Chandra, but also from other regions in the SNR where the magnetic field may be weaker. Similar conclusions are reported for Vela Junior supernova remnant even though in this case the hadronic scenario can not be ruled out. However, being of hadronic or leptonic origin, the GeV-TeV gamma-ray detections imply a low maximal energy for the accelerated particles of $\sim 100$~TeV, well below the knee of the cosmic-ray spectrum.

\subsection{Historical supernova remnants}
Two historical SNRs have been detected both at GeV and TeV energies: Cassiopeia A (Cas A) [\refcite{fermi_casa}, \refcite{magic_casa}, \refcite{veritas_casa}] and Tycho [\refcite{fermi_tycho}, \refcite{veritas_tycho}]. \\
Cas~A is the remnant of SN 1680. It is the brightest radio source in our Galaxy and its overall brightness across the electromagnetic spectrum makes it a unique laboratory for studying high-energy phenomena in SNRs. A multiwavelength modeling of Cas A does not allow a discrimination between the hadronic and leptonic scenarios. However, regardless of the origin of the observed gamma rays, this modeling implies that the total content of CRs accelerated in Cas~A is $\sim$(1 -- 2)$\times 10^{49}$ erg, and the magnetic field amplified at the shock can be constrained as B $\approx$ 0.12~mG. Even though Cas A is considered to have entered the Sedov phase, the total amount of CRs accelerated in the remnant constitutes only a minor fraction ($\sim2$\%) of the total kinetic energy of the supernova, which is well below the $\sim 10$\% commonly used to maintain the cosmic-ray energy density in the Galaxy.\\ 
Tycho's SNR (SN 1572) is classified as a Type Ia (thermonuclear explosion of a white dwarf) based on observations of the light-echo spectrum. Thanks to the large amount of data available at various
wave bands, this remnant can be considered one of the most promising object where to test the shock acceleration theory and
hence the CR -- SNR connection. First, using the precise radio and X-ray observations of this SNR, Morlino \& Caprioli (2011) [\refcite{morlino}] have shown that the magnetic field at the shock has to be $> 200 \mu$G to reproduce the data. Then, using  multiwavenlength data, especially the GeV and TeV detections, they could infer that the gamma-ray emission detected from Tycho cannot be of leptonic origin, but has to be due to accelerated protons (this result is consistent with another modeling proposed in \refcite{fermi_tycho}). These protons are accelerated up to energies as large as $\sim$500 TeV, with a total energy converted into CRs estimated to be about 12 per cent of the forward shock bulk kinetic energy. This is much more reasonable in the context of acceleration of Galactic cosmic-rays in SNRs.

\section{Where are the PeVatrons ?}
The recent GeV and TeV detections of supernova remnants confirm the theoretical predictions that supernova remnants can operate as powerful cosmic ray accelerators. However, if these objects are responsible for the bulk of galactic
cosmic rays, they should be able to accelerate protons and nuclei at least up to $10^{15}$~eV and therefore act as PeVatrons. Gabici and Aharonian (2007) [\refcite{gabici}] have shown that the spectrum of nonthermal particles extends to PeV energies only during a relatively short period of the evolution of the remnant since high energy particles are the first to escape from the supernova remnant shock. For this reason one may expect spectra of secondary gamma-rays extending to energies beyond 10 TeV only from less than 1~kyr old supernova remnants. In this respect, Tycho could be considered as a half-PeVatron at least, since 
there is no evidence of a cut-off in the VERITAS data. One may wonder how many PeVatrons are expected to be detectable in our Galaxy. A simple estimate has been provided by Gabici and Aharonian (2007): assuming a rate of $\sim$3 supernovae per century in our Galaxy, this directly implies that only a dozen of PeVatrons are present in the Galaxy on average and hence that they are likely to be distant and weak. This emphasizes the importance of TeV observations by the future generation of Cherenkov telescopes such as the Cherenkov Telescope Array (CTA) which will have a better effective area in the energy range already covered but that will also allow the observation up to 100 TeV of sources such as Tycho, therefore constraining the maximal energy at which protons are being accelerated in young SNRs.

\section*{Acknowledgements}
I thank all the members of the Fermi GALACTIC and HESS SNR-PWN working groups for valuable discussion. I gratefully acknowledge funding from the European Community (contract ERC-StG-259391).

\bibliographystyle{ws-procs9x6}
\bibliography{lemoine_snr}

\end{document}